\documentclass[aps,pre,preprint,showpacs]{revtex4-1}
\usepackage{ifpdf}
\usepackage{pdfsync}
\ifpdf
\usepackage{hyperref}
\else
\usepackage[hypertex]{hyperref}
\fi
\usepackage{graphicx,subfigure}
\usepackage{amsmath,amsfonts,amssymb}

\newcommand{\f}{\frac}

\newcommand{\intl}{\int\limits}

\newcommand{\eql}[1]{Eq.~\eqref{#1}}

\begin{document}
\title{The Vliegenthart-Lekkerkerker relation.
The case of the $Mie$-fluids}
%
\author{V.L. Kulinskii}
\email{kulinskij@onu.edu.ua}
\affiliation{Department for Theoretical
Physics, Odessa National University, Dvoryanskaya 2, 65026 Odessa, Ukraine}
\begin{abstract}
The Vliegenthart-Lekkerkerker
relation for the second virial coefficient
value at the critical temperature found in
[G. A. Vliegenthart and H. N. W. Lekkerkerker, J. Chem. Phys.
\textbf{112} 5364 (2000)] is discussed in connection with the scale invariant mean-field approach proposed in
[V.~L. Kulinskii and L.~A. Bulavin, J. Chem. Phys. \textbf{133} 134101
(2010)]. We study the case of the Mie-class potentials which is widely
used in simulations of the phase equilibrium of the fluids.
It is shown that due to the homogeneity property
of the $Mie$-class potentials it is possible to connect
the loci of the fluids with these model potentials
in different dimensions.
\end{abstract}
 \maketitle
\section{Introduction}\label{sec_intro}
Thermodynamic principle of the corresponding states (PCS)
\cite{pcs_guggenheim_jcp1945,book_prigozhisolut}
is one of the most vivid examples of the unifying nature of the
scale invariance. From the microscopic point of view
the PCS is based on simple
scaling properties of the interaction potential
which must be conformal
\cite{eos_pcsextended_jcp2000}.
Actually, the microscopic interactions in real
substances are more complex and do not conform with the
conditions at which the PCS can be derived rigorously from the
first principles of statistical mechanics.
To expand the range of the applicability of the PCS
its simple scaling form was extended to
include more parameters
which are connected with the most important properties of the
interparticle interactions of the simple fluids
(see e.g. \cite{liq_pcspitzer1_jamchemsoc1955}). The locus of the
critical point and corresponding compressibility factor
$Z_c = P_c/n_c\,T_c$, where $P_c,n_c,T_c$ are the critical
pressure, density and the temperature correspondingly,
is the core element of the PCS came originally
from van der Waals.

The interaction between nearest neighbors
dominates in a dense, condensed phase of the molecular system.
The liquid-gas critical point (CP) is adjacent to the
region of condensed liquid phase. Therefore its locus is
determined by the short scale properties of the interactions.
They are commonly the distance which corresponds to the minimum
of the potential
and the depth the potential well $\varepsilon$. But it is obvious
that the very existence of the CP is due to the attraction at
long distances \cite{eos_vliegenthartlekkerkerker_jcp2000,
eos_gasliqstability_prl1994,*eos_ljmetastable_physa1999}.
For the potential $\Phi(r)$ of the general
type its attractive and repulsive parts are independent.
The short-ranged characteristics of the potential weakly
depend on the long-range asymptotic behavior of the attractive part.
For potentials
of specific classes with simple analytical structure, e.g. the
conformal potentials, it is
possible to relate the short- and the long-ranged characteristics.
In such case, as it was noted in \cite{crit_scalingsmirnov_ufn2001en},
it is possible to construct the version of the PCS on the basis of a
long-range interaction only. The coordinates of the
critical point can be related with the long-range properties of the
potential. For the potentials of the Lennard-Jones type such scaling
approach was proposed in \cite{eos_zenomeglobal_jcp2010}.
The generalized
Lennard-Jones (LJ) potentials or Mie potentials
\cite{eos_miepotential_annphysik1903}:
\begin{equation}\label{miemn}
  \Phi(r;m,q) = \varepsilon\,\Phi_0(q/m)\,
  \left(\,\left(\,\f{\sigma}{r}\,\right)^q -
  \left(\,\f{\sigma}{r}\,\right)^{m}\,\right)\,,\quad q> m\,.
\end{equation}
serves as the example of conformal potentials. Here $\varepsilon$ is
the depth of the potential well,
the amplitude coefficient $\Phi_0(q,m)$ is as following:
\begin{equation}\label{mie_phi0}
  \Phi_0(q/m) = \f{m}{q-m}\,
  \left(\, \f{q}{m}\,\right)^{q/(q-m)}\,.
\end{equation}
Below we use common dimensionless units for the temperature
$T\to T/\varepsilon $ and the density $n \to n\,\sigma^d$,
where $d$ is the dimension.
Due to advances in computer simulations there is a great body of
results on the liquid-gas equilibrium
for the systems with the potentials \eqref{miemn}
\cite{crit_liqvamiepotent_jcp2000,eos_lj3d_pre2006,*eos_ljmn_jcp2008,*crit_ljmn_physleta2008}.

Obvious fundamental characteristic
which incorporates both repulsive and attractive
parts of the potential is the second
virial coefficient $B_2(T)$ \cite{book_hansenmcdonald}:
\begin{equation}\label{b2_3}
  B_{2}(T) = 2\,\pi\,\intl_{0}^{\infty}\,
    \left(\,1-e^{-\Phi(r)/T}\,\right)\,r^{2}\,dr\,.
\end{equation}
In Ref.~\cite{eos_vliegenthartlekkerkerker_jcp2000}
it was demonstrated that for many three
dimensional (3D) systems, where particles interact via
spherically symmetrical
pair interaction $\Phi(r)$, the relation:
\begin{equation}\label{b2virial6lekker}
  B_2(T_c)/v_0 = -C\,\,,
\end{equation}
holds, where $C$ is some constant. It should be noted that $C$ depends on
the number of dimensions $d$ too.
As it follows from \cite{eos_vliegenthartlekkerkerker_jcp2000}
the value of $C$ weakly depends
on the details of the potential as far as it
belongs to some class of potential functions.
In particular the authors of
Ref.~\cite{eos_vliegenthartlekkerkerker_jcp2000}
established that \emph{whereas the critical temperature
$T_c$ drops considerably upon narrowing the range of attraction,
the combination $B_2(T_c)/v_0$ remains practically constant}
\cite{eos_vliegenthartlekkerkerker_jcp2000}.
According to \cite{eos_vliegenthartlekkerkerker_jcp2000}
for the generalized potentials $\Phi(r;6,q)$, $C\approx 6.2$
in three dimensions. These results
put the question about the status of
\eqref{b2virial6lekker} as the mean of
parametrization the classes of potentials
in a sense more general than simple PCS.
In such general statement the problem seems rather difficult.
Nevertheless the possibility for its simplification may be found
in reduction this problem for the continuous systems to their lattice
analogues \cite{eos_latticecontinuum_jcp2000}.

In this paper we propose the derivation of the
Vliegerthart-Lekkerkerker relation for the potentials
of Mie-class \eqref{miemn} based on the global isomorphism between
the Lennard-Jones fluids and the lattice gas model
\cite{eos_zenomeglobal_jcp2010}.
The potentials \eqref{miemn} have simple analytical
structure and possess homogeneity dependence on the relevant
parameters $(m,q)$ (see Eq.~\eqref{mie_phi0}).

The structure of the paper is organized as
follows. In Section~\ref{sec_gliso} we discuss the dependence of
the coordinates of the CP on the parameters of the potential based
on the results of \cite{eos_zenomeglobal_jcp2010,eos_zenogenpcs_jcp2010}.
In Section~\ref{sec_lekker} we consider the dependence of the
Vliegerthart-Lekkerkerker parameter $C$ on the characteristics of
the Mie-class in $d$-dimensional case. The comparison with the known
results is given and some predictions are made which can be tested in
simulations. The discussion of the obtained results is in conclusive section.

\section{The locus of the critical point within the
global isomorphism approach}\label{sec_gliso}
It seems that there is no special symmetry property which
governs the locus
of the critical point in the continuum fluid.
Because of the regular structure and the
particle-hole symmetry of the configurations the lattice models
possess additional
symmetry properties which allow to calculate the locus of the
critical point exactly without the direct
calculation of the thermodynamic potential \cite{book_baxterexact}.
For the continuum fluid the particle-hole symmetry is broken and the binodal
is asymmetrical. The well-known rectilinear diameter law gives the
representation of this asymmetry \cite{pcs_guggenheim_jcp1945,crit_diam0,*crit_diam1_young_philmag1900,
*crit_diambenzene_physrev1900}.
Recently the relation between the phase diagrams
of the lattice gas and the simple liquid have been proposed in
\cite{eos_zenome_jphyschemb2010}. It is based on
the mapping between the fluid part of the phase diagram of the
Lennard-Jones (LJ) fluids in the form of the projective
transformation:
\begin{equation}\label{projtransfr_nx}
  n =\, n_*\,\f{x}{1+z \,t}\,,\quad
  T =\, T_*\,\f{z\, t}{1+z \,t}\,,
\end{equation}
where $x,t$ are the density and temperature variables of the
lattice gas, $n$ and $T$ are the corresponding quantities for
the continuum fluid. Here $n_*$ and $T_*$ are the
parameters which will be defined below (see Eq.~\eqref{cp_fluid}).
The temperature variable $t$ is defined so
that its value at the critical point (CP) is
$t_c  = 1$. Here the parameter $z$ is determined by the correspondence
between the loci of critical points of fluid $(n_c,T_c)$ and
the lattice gas $x_c=1/2\,, t_c=1$:
\begin{equation}\label{z_c}
  z = \f{T_c}{T_{*} - T_c}\,.
\end{equation}
The inverse transformation has the form:
\begin{equation}\label{projtransfr_nx_inv}
x =   \f{n}{1-T/T_*} \,,\quad
t = \f{1}{z}\,\f{T}{T_*-T} \,.
\end{equation}
The main assumption which governs the simple form of the transformation
\eqref{projtransfr_nx} is the validity
of the rectilinear diameter law for the density in a broad
temperature interval
of liquid-gas equilibrium \cite{pcs_guggenheim_jcp1945}.
The inverse form \eqref{projtransfr_nx_inv}
gives the procedure of the symmetrization
of the binodal of the fluid in terms of the variable
$x$ which is combination of the density $n$
and the temperature $T$ of the continuum fluid.
The idea that the difference between irregularity
configuration for continuum fluids and regularity of
configurations of lattice models
is unimportant for order-disorder transitions
was pronounced by K.S. Pitzer in \cite{eos_pitzer_purapplchem1989}.
It was stressed that the main difference is the shape of the
holes in real or continuum liquid and that in the lattice gas.
This is the reason why the behavior of continuum liquids
beyond the fluctuational region is different from that of the
lattice gas (Ising model). The global character of the
transformation \eqref{projtransfr_nx}
shows that the particle-hole simplified picture is valid not only
in near critical region but also far away from it. This rehabilitates
the hole theory for expanded liquids \cite{liq_barkerhenderson_rmp1973}
which gives the possibility to derive \eqref{projtransfr_nx} from
the microscopic point of view. Yet the transformation
Eq.~\eqref{projtransfr_nx} is not exact since its simple form is
heavily based on the validity of the law of the rectilinear diameter.
In the close vicinity of
the critical point there are singular fluctuation corrections to
the classical law of the rectilinear diameter
\cite{crit_rehrmermin_pra1973,*crit_aniswangasymmetry_pre2007,*crit_can_diamsing_kulimalo_physa2009}.
Nevertheless, it is valid for the Lennard-Jones fluids in a broad
temperature interval and is widely used to
locate the critical point \cite{book_frenkelsimul}.
The numerical data of
\cite{eos_yukawa_jcp2007,crit_longrangecampatey_jcp2001,*crit_supercrit_physb2001,*eos_ljfluid_jcp2005}
are consistent with this law and therefore simple
form of Eq.~\eqref{projtransfr_nx}
is applicable for these systems. The results of
\cite{crit_globalisome_jcp2010} for the mapping between
Lennard-Jones fluids
and the Ising model in 2 and 3 dimensions support
the validity of such approximation.

From Eq.~\eqref{projtransfr_nx} the loci of the CP for fluid
and lattice gas are connected by simple relations:
\begin{equation}\label{cp_fluid}
  n_{c} = \f{n_*}{2\left(\,1+z\,\right)}\,,\quad
  T_{c} = T_{*}\, \f{z}{1+z}\,,
\end{equation}
where the parameter $n_*$ is given by:
\begin{equation}\label{nbtb}
n_*= T_*  \,\f{B'_2\left(\,T_*\,\right)}
{B_3\left(\,T_*\,\right)}\,,
\end{equation}
and $T_*$ is equal to the Boyle temperature in the vdW
approximation $T^{(vdW)}_{B}= a/b$ \cite{eos_zenomeglobal_jcp2010,crit_globalisome_jcp2010}.
Here $b = 4\, v_0$ is the fourfold of the
eigenvolume of the particles $v_0 = \pi \sigma^3/6$ and
\begin{equation}\label{a_vdw}
  a = -2\pi\,\intl_{\sigma}^{+\infty}\Phi_{\text{attr}}(r)\,r^2\,dr\,\,,
\end{equation}
where $\Phi_{\text{attr}}(r)$ is the attractive part of
the potential $\Phi(r)$.
As has been shown in
\cite{eos_zenogenpcs_jcp2010} it is possible to relate the
parameter $z$ of the transformation \eqref{projtransfr_nx}
with the exponent of the attractive part of the potential
$\Phi_{\text{attr}}(r)\simeq -r^{-m}\,$. It is based on
the relation
\begin{equation}\label{cp_scalingalpha}
- \f{d\,\ln{\left(\,T_c/T_{*}\,\right)}}
{d\,\ln{\left(\,n_c/n_{*}\,\right)}} = \f{1}{z}\,\,,
\end{equation}
which directly follows from Eq.~\eqref{cp_fluid} and
the scaling relation
\begin{equation}\label{tcphi}
  \f{d\,\ln \left(\,T_c/T_{*} \,\right) }{d\,\ln n_c/n_{*}}
  =
  \f{d\,\ln \left(\,\Phi_{\text{attr}}(n^{-1/d}_c)/T_{*}\,\right) }
  {d\,\ln n_c/n_{*}}\,\,,
\end{equation}
between the critical temperature and
that if $\Phi_{\text{attr}}(r)\simeq -r^{-m}\,$. This relation
generalizes commonly used reasoning that the critical
temperature $T_c$ is of order of the magnitude of the
potential well \cite{crit_scalingsmirnov_ufn2001en}.
Then in $d$-dimensions $z$  is determined as:
\begin{equation}\label{zdm}
  z = \f{d}{m}\,.
\end{equation}
According to Ref.~\cite{eos_vliegenthartlekkerkerker_jcp2000}
$C\approx 6.3$ in $3D$
for the potentials of the LJ type $\Phi(r;6,q), q\ge 7$
as well as for $\Phi(r;m,2m), m\ge 6$.
In Ref.~\cite{eos_zenogenpcs_jcp2010} it was demonstrated
that the results of Eq.~\eqref{cp_fluid} for the critical
temperature are very close to those following from
Eq.~\eqref{b2virial6lekker} for the potentials with $r^{-6}$
asymptotic of the attractive part.

If the interaction does not have the hard core,
there is an arbitrariness in the definition of the
diameter of the particle $\sigma$ in Eq.~\eqref{a_vdw}.
Usually, it is chosen
as the root of the equation $\Phi(\sigma ) = 0$.
The locus of the CP determined by \eql{cp_fluid}
depends on the definition of
soft core diameter $\sigma$ and changes with the
change of this scale.
Nevertheless,
it should be noted that Eq.~\eqref{b2virial6lekker} does
not depend on the spatial scale used to determine the size
of the molecule like $\sigma$. This size is introduced
from the physical reasonings to distinguish between
the short and the long range part of the potential (see e.g.
\cite{liq_barkerhendersondiam_jcp1967,*liq_weekschandler_jcp1971,
*liq_bhdiameter_jcp2002,eos_ljsimplefluids_jcp2006}).

Note that according to Eq.~\eqref{miemn} the quantity $\Phi_0$
is the homogeneous function of $m$ and $q$. The change of the
the scale for $\sigma$ leads to simple rescaling of the parameter $T_*$
because it is determined by the
long-range power-like asymptotic behavior of the interaction.
This points to the connection between \eqref{b2virial6lekker} and
\eqref{cp_fluid}.
Moreover, Eq.~\eqref{cp_fluid} in principle gives the value of
the critical density too, while
Vliegenthart-Lekkerkerker relation
\eqref{b2virial6lekker} as it is formulated originally
does not allow to estimate $n_c$. Therefore
it is interesting to clarify the interrelation
between Eq.~\eqref{cp_fluid} and Eq.~\eqref{b2virial6lekker}.
We consider the results of
Ref.~\cite{eos_vliegenthartlekkerkerker_jcp2000} for the generalized
Lennard-Jones potentials as the
nontrivial extension
of the PCS. The class of PCS consists of the family of
potential functions given by Eq.~\eqref{miemn}. This inference
is supported by the computer simulations
\cite{crit_liqvamiepotent_jcp2000,crit_ljmn_physleta2008}.

Below we elucidate the nature
of  the Vliegenthart-Lekkerkerker relation \eqref{b2virial6lekker}
and its connection with the global isomorphism approach.
We use the results of
Ref.~\cite{eos_zenogenpcs_jcp2010}
to connect the value $C$ with the parameter $z$.
The latter represents
the class of the thermodynamic similarity in
accordance with the
generalized principle of corresponding states
\cite{eos_zenogenpcs_jcp2010}.

\section{The nature of the
Vliegenthart-Lekkerkerker relation within the global isomorphism approach}
\label{sec_lekker}%
The relation Eq.~\eqref{b2virial6lekker} looks quite
unusual in a sense that it states
the relation between the characteristic $B_2$ determined
in low density region $n\to 0$
with $T_c$. The last is related with the moderate densities
of liquid state. In the framework of the approach of
Ref.~\cite{eos_zenome_jphyschemb2010}
this is naturally explained since the
Boyle temperature $T^{(vdW)}_{B} = T_*$
is connected with the critical one $T_c$
according to Eq.~\eqref{cp_fluid}. In previous works
\cite{eos_zenogenpcs_jcp2010,crit_globalisome_jcp2010}
we show how the conception of the global isomorphism proposed in
Ref.~\cite{eos_zenome_jphyschemb2010} can be applied to formulate the
scale invariant mean-field approach for calculation of the locus
of the CP.

We are not restricted  by the 3D case and consider
the case of general
dimension $d>1$ so that the second virial coefficient
is as following:
\begin{equation}\label{b2_d}
  B_{2}(T) = \f{S_{d}}{2}\,\intl_{0}^{\infty}\,
    \left(\,1-e^{-\Phi(r)/T}\,\right)\,r^{d-1}\,dr\,,
\end{equation}
where $S_{d} = \pi^{d/2}/\Gamma\left(\,\f{d}{2}\,\right)$
is the unit sphere in $d$-dimensional space.
This gives us the
possibility to establish the dependence of $C$ on the
relevant parameters in
$d$-dimensional case.

For the temperature we will use the common unit, the absolute
value of the minimum $\varepsilon$ of the potential well.
The spatial scale is commonly connected with the parameter $\sigma$.
But in accordance with the scale invariant nature of
Eq.~\eqref{b2virial6lekker} one can choose it in appropriate way.
We use such freedom for the definition of the spatial scale
which separate the intervals of distance where the repulsive
and attractive forces dominate correspondingly.

It is easy to derive the estimate for the parameter $C$ using the high-temperature
asymptotic expansion for Eq.~\eqref{b2_d} (see e.g. \cite{book_ll5_en}):
\begin{equation}\label{b2virial_vdw}
    B_2(T) = b \left(1-T_*/T\right)+ o\left(\,T_*/T\,\right)\,\,.
\end{equation}
Taking into account that in $d$ dimensions $b=2^{d-1}\,v_0$
and using the relation \eqref{cp_fluid}, from
Eq.~\eqref{b2virial6lekker} we obtain:
\begin{equation}\label{c_hightempasypt}
  C\approx \f{2^{\,d-1}}{z}\,.
\end{equation}
For the LJ potential $\Phi(r;6,12)$ in $3D$ case for which $z=1/2$ this gives
$C\approx 8$, for $2D$ case $z = 1/3$,  $C\approx 6$ correspondingly.

Below we improve the estimate \eqref{c_hightempasypt} for $C$
in a case of $Mie$ potentials \eqref{miemn}. For these potentials the left
hand side of Eq.~\eqref{b2_d}
transforms into:
\begin{equation}\label{vl_scaleinv_me}
 B_2(T)/v_0 = 2^{\,d-1}\,b_0\,,
\end{equation}
where
\begin{equation}\label{b0}
b_0(T/\Phi_0,m/d,q/m) = \intl_{0}^{\infty}\,
    \left(\,1-\exp\left(\,\f{\Phi_{0}/T}{x^{m/d}}
    \left(1 -x^{-(q-m)/d}\,\right)\,\right)\,\right)\,dx\,,
\end{equation}
and $x = r^d$ and $v_0 = S_d/d\,
\left(\,\sigma/2 \,\right)^d$ is the $d$-dimensional volume.
Notably,
the parameter $b_0$ depends on its arguments homogeneously.
So if the critical temperature $T_c$ is determined by some
constraint for $B_{2}(T)$ with the dimension $d$ fixed,
then it is the homogeneous function of
the relevant parameters $(m/d,q/m)$. We can say that the last represents the corresponding homogeneity class. The same homogeneous
dependence is inherent to the
parameter $T_*$ for the potentials of
$Mie$-type \eqref{miemn}:
\begin{equation}\label{t_bvdw_d}
T_{*} = -d\,\intl^{+\infty}_{1}
\Phi_{\text{attr}}(r)\,\,r^{d-1}\,dr\, =
\frac{\left(\frac{q}{m}\right)^{\frac{q}{q-m}}}{(m/d-1) (q/m-1)}\,.
\end{equation}
Therefore, $T_{*}$ is the same for the potentials
with the same ratio $q/m$. This allows to use simple scaling
considerations to connect the critical temperatures for the systems
with different potentials belonging to the same homogeneity class.
Besides, in view of the isomorphism
with the lattice gas we can state that
\begin{equation}\label{td1td2}
T_c(d_1)/T_c(d_2)\approx d_1/d_2\,,
\end{equation}
for the potentials of the same
homogeneity class $(m/d,q/m)$. Indeed, the expression \eqref{cp_fluid}
assumes that the critical temperature of the isomorphic
lattice model is set to unit $t_c = 1$.  So the correct
comparison of the critical temperatures for the
potentials in different dimensions demands the corresponding
scaling since $t_c$ is proportional to the number
of the nearest neighbors, which in its turn proportional
to the dimension of the lattice $d$. Of course, the relation
\eqref{td1td2} is approximate since the fluctuations are
neglected. Nevertheless it is instructive to check the
validity of Eq.~\eqref{td1td2} and the statement that the
critical temperatures for the potentials
belonging to the same similarity class can be connected via
simple scaling relations. E.g. the potentials $\Phi(r;6,12)$ in
$d=2$ and $\Phi(r;9,18)$ in $d=3$ fall into the same homogeneity
class $(2,2)$. Therefore
\[T_c(d=3; 9,18)/T_c(d=2; 6,12) \approx 3/2\,.\]
Though the Vliegenthart-Lekkerkerker relation itself
does not allow
to calculate other parameters of the critical point like
the critical density and the pressure, they can be calculated
using the methods developed in \cite{eos_zenogenpcs_jcp2010}.
The value of the critical density is calculated in accordance
with Eq.~\eqref{cp_fluid}.
Let us check \eqref{td1td2} considering
the potentials $\Phi(r;m,2m)$ for $d=2,3,4$. For these potentials the
parameter $\Phi_0$ takes the constant value $\Phi_0 = 4$ so that:
\begin{equation}\label{bn2n}
  B_{2}(T) = 2^{d-1}\,b_0(T/4;m/d,2)\,.
\end{equation}
The results of calculations are presented in
Tables~\ref{tab_tc136}-\ref{tab_tc236} along with the
known numerical estimates. We take the results for the critical
temperature of the
Lennard-Jones potential $\Phi(r;6,12)$ as known since they can be
extracted from the corresponding lattice models (see also
\cite{eos_zenomeglobal_jcp2010,crit_globalisome_jcp2010}).
\begin{table}
  \centering
  \begin{tabular}{|c|c|c|c|}
\hline
    $d$ & 2, $\Phi(r;6,12)$ & 3, $\Phi(r;9,18)$ & 4, $\Phi(r;12,24)$\\
\hline
    $T_c$ & \hphantom{0.5}0.5\hphantom{0.5} &
    \hphantom{0.5}0.75 \hphantom{0.5}
    & \hphantom{0.5}1.\hphantom{0.5}\\
\hline
    $T^{(num)}_c$ &\hphantom{0.5}0.515, \cite{crit_lj2dim_jcp1991}
    \hphantom{0.5} &
    \hphantom{0.5}0.73\,, \cite{eos_vliegenthartlekkerkerker_jcp2000} \hphantom{0.5}&
    \hphantom{0.5} ?  \hphantom{0.5}\\
    \hline
        $n_c$ & \hphantom{0.5}0.353\hphantom{0.5} &
    \hphantom{0.5}0.436 \hphantom{0.5}
    & \hphantom{0.5}0.55\hphantom{0.5}\\
\hline
    $n^{(num)}_c$ & \hphantom{0.5}0.355, \cite{crit_lj2dim_jcp1991} \hphantom{0.5} &
    \hphantom{0.5}0.354, \cite{eos_vliegenthartlekkerkerker_jcp2000} \hphantom{0.5}
    & \hphantom{0.5}?\hphantom{0.5}\\
\hline
  \end{tabular}
  \caption{The critical temperatures for the
$Mie$  potentials of the similarity class
  $(m/d,q/m) = (3,2)$. It includes $\Phi(r;6,12)$ in
  $d=2$, $\Phi(r;9,18)$ in $d=3$ and $\Phi(r;12,24)$ in $d=4$.}\label{tab_tc136}
\end{table}
\begin{table}
  \centering
  \begin{tabular}{|c|c|c|c|}
\hline
    $d$ & 2, $\Phi(r;4,8)$ & 3, $\Phi(r;6,12)$ & 4, $\Phi(r;8,16)$\\
\hline
    $T_c$ & \hphantom{0.5}0.89\hphantom{0.5} &
    \hphantom{0.5}1.33 \hphantom{0.5}
    & \hphantom{0.5}1.78\hphantom{0.5}\\
\hline
    $T^{(num)}_c$ &\hphantom{0.5}?\hphantom{0.5} &
    \hphantom{0.5}1.313\,, \cite{crit_liqvamiepotent_jcp2000}
    \hphantom{0.5}&
    \hphantom{0.5} ?  \hphantom{0.5}\\
    \hline
       $n_c$ & \hphantom{0.5}0.283\hphantom{0.5} &
    \hphantom{0.5}0.322\hphantom{0.5}
    & \hphantom{0.5}0.392\hphantom{0.5}\\
\hline
  $n^{(num)}_c$ & \hphantom{0.5}?\hphantom{0.5} &
    \hphantom{0.5}0.316, \cite{crit_liqvamiepotent_jcp2000}  \hphantom{0.5}
    & \hphantom{0.5}?\hphantom{0.5}\\
\hline

  \end{tabular}
  \caption{The critical temperatures for the
$Mie$  potentials of the similarity class
  $(m/d,q/m) = (2,2)$.  It includes $\Phi(r;4,8)$ in
  $d=2$, $\Phi(r;6,12)$ in $d=3$ and $\Phi(r;8,16)$ in $d=4$.}\label{tab_tc248}
\end{table}
\begin{table}[hbt!]
  \centering
  \begin{tabular}{|c|c|c|c|}
\hline
    $d$ & 2, $\Phi(r;3,6)$ & 3, $\Phi(r;9/2,9)$ & 4, $\Phi(r;6,12)$\\
\hline
    $T_c$ & \hphantom{0.5}1.6\hphantom{0.5} &
    \hphantom{0.5}2.4 \hphantom{0.5}
    & \hphantom{0.5}3.2\hphantom{0.5}\\
\hline
    $T^{(num)}_c$ &\hphantom{0.5}?\hphantom{0.5} &
    \hphantom{0.5}?\hphantom{0.5}&
    \hphantom{0.5} 3.4\,, \cite{crit_lj4dim_jcp1999}  \hphantom{0.5}\\
    \hline
      $n_c$ & \hphantom{0.5}0.333\hphantom{0.5} &
    \hphantom{0.5}0.353\hphantom{0.5}
    & \hphantom{0.5}0.41
    \footnote{In \cite{eos_zenomeglobal_jcp2010} we gave another value
    $n_c \approx 0.404$. The difference is caused by different
    integration method and the precision for the calculation
 of multidimensional integral for $B_3$ in higher dimensions ($d\ge 4$).
 Here we obtained $n_c \approx 0.408$ and rounded the result up
 to hundredth.}\hphantom{0.5}\\
\hline
  $n^{(num)}_c$ & \hphantom{0.5}?\hphantom{0.5} &
    \hphantom{0.5}? \hphantom{0.5}
    & \hphantom{0.5}0.34, \cite{crit_lj4dim_jcp1999}\hphantom{0.5}\\
\hline
  \end{tabular}
  \caption{The critical temperatures for the
$Mie$  potentials of the similarity class
  $(m/d,q/m) = (3/2,2)$.  It includes $\Phi(r;3,6)$ in
  $d=2$, $\Phi(r;9/2,9)$ in $d=3$ and $\Phi(r;6,12)$ in $d=4$.}
  \label{tab_tc236}
\end{table}
%
%

Using relations \eqref{cp_fluid} and \eqref{zdm}
for the case of the LJ potential $\Phi(r;6,12)$
we obtain:
\begin{equation}\label{b3612}
  b_0(T_c/4;2,2) = 3^{1/4}\, \vphantom{F}_1F_1
  \left(-\frac{1}{4};\frac{1}{2};\frac{3}{4}\right) \Gamma
   \left(\frac{3}{4}\right)-3^{3/4}
     \vphantom{F}_1F_1\left(\frac{1}{4};\frac{3}{2};
     \frac{3}{4}\right) \Gamma
   \left(\frac{5}{4}\right)\approx 1.51\,,
\end{equation}
for $d=3$. Here $\vphantom{F}_1F_1$ is the Kummer confluent
hypergeometric function
\cite{book_abramovitzstegun}. In accordance with
Eq.~\eqref{vl_scaleinv_me}
this gives  $C\approx 6.04$. In two-dimensional case for the
LJ potential $\Phi(r;6,12)$ the value of the
Vligerthart-Lekkerkerker constant $C$ is determined by
\begin{equation}\label{b2612}
b_0(T_c/4;3,2) =
\sqrt{2}\, \vphantom{F}_1F_1\left(-\frac{1}{6};\frac{1}{2};2\right)
\Gamma\left(\frac{5}{6}\right)-
2\, \vphantom{F}_1F_1\left(\frac{1}{3};\frac{3}{2};2\right) \Gamma
   \left(\frac{4}{3}\right)\approx 3.81\,,
\end{equation}
which leads to $C \approx 7.6$. Finally, for the homogeneity class $(3/2,2)$:
\begin{equation}\label{b4612}
b_0(T_c/4;3/2,2) = \left(\,\f{5}{4}\,\right)^{1/3}
\left(\vphantom{F}_{1}F_{1}\left(-\frac{1}{3};\frac{1}{2};\frac{5}{16}\right)
   \Gamma \left(\frac{2}{3}\right)-\sqrt{5}\,\Gamma\left(\frac{7}{6}\right)\,
   \vphantom{F}_{1}F_{1}\left(\frac{1}{6};\frac{3}{2};\frac{5}{16}\right)
   \right)
\approx 1.19\,,
\end{equation}
with the corresponding value $C \approx 9.5$.
This corrects the high-temperature
estimate \eqref{c_hightempasypt}.

\section{Conclusions}
In this paper we derive the
Vliegenthart-Lekkerkerker relation
for the value of the second
virial coefficient at the critical temperature
\cite{eos_vliegenthartlekkerkerker_jcp2000}
within the approach proposed in \cite{eos_zenogenpcs_jcp2010}
for the potentials of
$Mie$-class. The very existence of such relation has appeared
as the direct consequence of the global isomorphism between the
fluids with the simple conformal potentials and the lattice
models. We extended the applicability of this relation by
considering the general case of $d$ dimensions. The estimates for
the loci of the critical points of the systems with $\Phi(r;m,2m)$-type
potentials are obtained. These estimates can be checked
in simulations and it would be interesting to investigate the
relations between the critical parameters for the systems
with the $Mie$ potentials of the same homogeneity class
but with different dimensions. This provides the test for the validity
of the proposed approach.

The application of the proposed approach to other types of the
potentials with nonalgebraic behavior, like
hard spheres plus attractive Yukawa potential, Morse potential
or square-well potential is connected with the derivation of the
relation between the parameter $z$ of the transformation
\eqref{projtransfr_nx}
and the relevant integral characteristics of the potential.
E.g., the relation \eqref{tcphi} can not be applied
directly to the square well potential because of nonanalytic dependence
$\Phi(r)$. Obviously,
these characteristics should be related with the symmetry
properties of the potential under the scaling of its parameters,
namely the effective range of interaction. The
results of Ref.~\cite{eos_vliegenthartlekkerkerker_jcp2000} show that the
value of the Vliegenthart-Lekkerkerker parameter $C$ differs essentially
for Lennard-Jones like potentials $\Phi(r;6,q)\,, q \ge 7$ and
square-well potential. In the last case there is strong dependence
on the width of the potential well which governs the range of the
interaction. The weak dependence of $C$ on the exponent $q$
of the repulsive part for $\Phi(r;6,q)$ potentials in naturally explained
within the global isomorphism approach because of Eq.~\eqref{zdm} and
Eq.~\eqref{cp_fluid}. The square-well potential is the continuum version
of the nearest neighbor interaction in the lattice model. So
it is possible to connect the dependence of the critical parameters
for the square-well potential and the corresponding results for the
regular lattice gas model. These topics will be the subjects of
further studies.


%

\end{document}